# Influence of in-plane and bridging oxygen vacancies of $SnO_2$ nanostructures on $CH_4$ sensing at low operating temperatures

Venkataramana Bonu,[1*] A. Das,[1*] Arun K Prasad,[1] Nanda Gopala Krishna,[2] Sandip Dhara,[1] and A. K.Tyagi.[1]

[1]*Surface and Nanoscience Division, Indira Gandhi Center for Atomic Research, Kalpakkam-603102, India.*

[2]*Corrosion Science & Technology Group, Indira Gandhi Center for Atomic Research, Kalpakkam-603102, India.*

Role of 'O' defects in sensing pollutant with nanostructured $SnO_2$ is not well understood, especially at low temperatures. $SnO_2$ nanoparticles were grown by soft chemistry route followed by subsequent annealing treatment under specific conditions. Nanowires were grown by chemical vapor deposition technique. A systematic photoluminescence (PL) investigation of 'O' defects in $SnO_2$ nanostructures revealed a strong correlation between shallow donors created by the in-plane and the bridging 'O' vacancies and gas sensing at low temperatures. These $SnO_2$ nanostructures detected methane ($CH_4$), a reducing and green house gas at a low temperature of 50 °C. Response of $CH_4$ was found to be strongly dependent on surface defect in comparison to surface to volume ratio. Control over 'O' vacancies during the synthesis of $SnO_2$ nanomaterials, as supported by X-ray photoelectron spectroscopy and subsequent elucidation for low temperature sensing are demonstrated.

**Electronic mail:** dasa@igcar.gov.in, ramana9hcu@gmail.com.

Since two decades, research community has been working intensely for the development of metal oxide nanostructure based gas sensors for low temperature operation along with improved response, selectivity and stability.[1] In this context, sensor performances of widely used $SnO_2$ metal oxide have been improved by reducing size to nano dimension and also by incorporating additives.[2,3] However, apart from the size effect surface chemistry of metal oxides play a crucial role in gas sensing process.[4-6] In-plane oxygen ('O') vacancies are believed to be responsible for low temperature sensing of oxidizing $NO_2$ in case of $SnO_2$ nanocrystals.[4] The gas response was found to vary linearly with the amount of 'O' vacancy, which was supported by photoluminescence (PL) studies.[5] Apart from providing the large surface area and subsequent high response; the influence of nanocrystallite size on the low temperature gas sensing, especially for reducing gases like $CH_4$, CO, are yet to be understood. Recently, template growth of porous $SnO_2$ nanostructures using carbon nanotube responded even at room temperature (RT) against 1000 ppm $CH_4$ gas.[7] However, the mechanism of detection is not well understood.

$SnO_2$ is an intrinsic *n*-type wide band gap semiconductor. Oxygen vacancy related defects are known to make it intrinsically *n*-type semiconductors.[2] These defects also play crucial role in gas sensing by controlling electrical conduction process as a probe parameter. There are several reports on defect study of $SnO_2$ using PL and gas sensing related to size of nanostructures. However, role of defects in controlling low temperature sensing remains to be explored. Low temperature detection of stable reducing gas like $CH_4$ which is a colorless, odorless and highly flammable green house gas is an industrial demand as it constitutes a major ingredient of natural gas and being widely used in industrial and domestic appliances.[8]

In this context, we demonstrate the prevalent role of particular defects like in-plane and bridging 'O' vacancies in the temperature dependence of $CH_4$ sensing using a series of nanosized $SnO_2$ particles and nanowires (NWs). Systematic PL studies including temperature dependence were carried out to deduce the specific role of shallow donors, which were created by the in-plane and bridging 'O' vacancies of nanomaterials, in the sensing process.

$SnO_2$ nanoparticles (NPs) with varied size were synthesized by annealing pristine $SnO_2$ at 300 °C, 500 °C and 800 °C for 1 h in air atmosphere. Pristine $SnO_2$ was synthesized by soft chemical method from a reaction mixture of $NH_4OH$ (2M) and aqueous $SnCl_4$ (1M) solution under continuous magnetic stirring at 80 °C. The detailed synthesis process of the pristine sample was reported earlier.[9] Mixture of pristine $SnO_2$ and graphite powder (Alfa

Aesar, 99.9995%) in a 3:1 weight ratio was also used for the growth of NWs on 3 nm thick Au coated Si(100) substrate following chemical vapor deposition (CVD) technique at 1000 °C under 60 sccm Ar gas flow rate. Au was used as catalyst in the vapor liquid-solid (VLS) growth process.

Morphological studies of the synthesized materials were performed by field emission scanning electron microscopy (FESEM; Zeiss SUPRA 55) and high resolution transmission electron microscopy (HRTEM; Zeiss Libra 200). X-ray photoelectron spectroscopy (XPS; M/s SPECS GmbH, Germany) was employed to study the oxidation state of Sn and stoichiometry of the surface of NPs from the quantitative analysis of both Sn and O. PL (InVia, Renishaw) studies using excitation wavelength of 325 nm (He-Cd Laser) were carried out to investigate the defects in all the NPs and NWs. The study was performed with 2400 gr/mm grating for monochromatization area and thermoelectric cooled CCD detector in the backscattering configuration. Ultra high pure $CH_4$, $N_2$ and $O_2$ were used for the gas sensing experiments. Change in resistance of the sensor device upon continuous gas exposure was recorded using micro-Ohm meter (Agilent 34401). Details of the dynamic gas exposure facility were reported earlier.[10] Gas sensing devices were fabricated by dispersing NPs (1mg/ml) and NWs (0.5 mg/ml) using isopropanol solution on the interdigitated Au electrodes (IDE) by using a spin coater at 600 rotation per minute. The Au electrodes of 50 nm thickness were coated by DC sputtering (Vacuum Techniques (P) Ltd, India) on $SiO_2$ (300 nm)/Si substrates. The contact between Au electrode and $SnO_2$ nanostructures was found to be Ohmic. Gas exposure was carried out at various temperatures.[10]

HRTEM studies (Figs. 1(a) and 1(b)) reveal that the average sizes of spherical NPs annealed at 300 and 500 °C are 4 and 9 nm, respectively. FESEM image (Fig. 1(c)) of the NPs annealed at 800 °C shows spherical NPs of average size 25 nm. Densely packed NWs grown on the Si substrate are shown in the FESEM image (Fig. 1(d)) with length and diameter measuring around 100 μm and 150 nm, respectively. The most stable plane (110) of rutile tetragonal phase of $SnO_2$ was observed in all the zoomed HRTEM images in the insets (Fig. 1) with a $d$ spacing value of 3.34 Å (JCPDS card # 41-117). Au NP appearing as dark contrast at the tip of the NW (bottom inset in Fig. 1(d)) supported the VLS growth mechanism.[11]

XPS analysis for NPs of sizes of 4 and 25 nm were typically carried out to investigate the surface chemistry changes with annealing temperature. Figs. 2(a,b) and 2(c,d) show Sn3$d$ and O1$s$ spectral windows of 4 nm and 25 nm NPs, respectively. The $SnO_2$ appears as a spin-orbit doublet at 487.3 eV (3$d_{5/2}$) and 495.7 eV (3$d_{3/2}$) with an area ratio of 1.5 inferring the presence of $Sn^{4+}$ chemical state in the $SnO_2$ NPs.[12] Atomic weight percentage ratios between elements 'O' and 'Sn' (O:Sn) were approximately found to be 1.7 for 4 nm NPs and 1.5 for 25 nm NPs. Both values were found to be lower than that of the measured value of ~2 for pristine $SnO_2$. It indicates increasing oxygen deficiency as the nanoparticle grows in size. This increased defects density manifested in strong desorption of 'O's from the surface of NPs due to annealing of the as-prepared material at high temperatures of 500 and 800 °C. However, annealing at relatively low temperature of 300 °C allowed particle to grow 4 nm NPs with better stoichiometry than that for the NPs of larger size.

PL study was carried out for the identification of various 'O' defects in $SnO_2$. A broad PL peak around 2 eV (Fig. 3(a)) for all the samples was assigned to surface 'O' vacancies in the $SnO_2$ nanostructures.[4,13,14] PL intensity increases with increasing size of the NPs under same experimental conditions. Enhancement in the PL intensity points to the existence of increased amount of 'O' vacancies in the NPs as indicated in the XPS analysis. The broad PL peak was deconvoluted into Gaussian peaks centered at 1.81, 1.97, 2.12, 2.28 and 2.45 eV (Fig. S1). Temperature dependent PL studies provided further insight of the nature of defects and its relative position in the energy levels with respect to conduction band (CB) minimum in the band diagram. This study was carried out typically for 25 nm NPs (supplementary Fig. S2). Systematic changes for the intensities of the peaks at 1.97, 2.12, 2.28 and 2.45 eV were observed with respect to the temperature (Fig. S3). This result indicates that all the four defect states belonging close to the CB follow a sequence for shallow donors (SD) as SD1 (2.45 eV) > SD2 (2.28) > SD3 (2.12) > SD4 (1.97). It is now shown schematically in the band diagram (Fig. 3(b)). The density functional theory (DFT) calculation based on all electron Gaussian approximation estimated that an acceptor state just ~1 eV above the valance band (VB) could appear due to the stable 'O' vacancies.[15] DFT calculations using generalized gradient approximation (GGA) also predicted luminescence transitions around 2 and 2.4 eV which were attributed to the bridging and the in-plane 'O' vacancies, respectively.[4] By using similar calculations Liu et al.[16] obtained density of states for the bridging and the in-plane 'O' vacancies. Their calculations revealed that the in-plane 'O' vacancies were responsible for shallow donor states while the bridging 'O' vacancies were to form relatively deeper states. D. Dutta et al.[17] reported luminescence at ~ 2.4 eV for $SnO_2$ quantum dots due to singly charged in-plane 'O' vacancies which was probed independently using electron spin resonance technique. In accordance to above discussion, SD1 and SD2 were assigned to singly charged in-plane vacancies while SD3 and SD4 were correlated to bridging 'O' vacancies (Fig. 3(b)).

In fact, the PL intensities of different shallow donors were also found to vary in various nanosized $SnO_2$ NPs. For instance, the intensities of peaks at 2.45 eV (SD1) and 2.3 eV (SD2) were observed to contribute strongly for the 9 nm NPs in comparison to the 4 nm NPs. However, 4 nm $SnO_2$ NPs exhibited stronger PL peak at 1.97 eV (SD4) than those for 2.3 eV (SD2) and 2.45 eV (SD1) (Fig S1a). In case of 9 nm NPs, the ratio between total area under peaks 1.97 eV (SD4) and 2.12 eV (SD3), related to the bridging 'O' vacancies and total area under PL peaks of 2.3 (SD2) and 2.45 eV (SD1) contributing to the in-plane vacancies was 8:5. In contrast, the ratio was 5:9 for the $SnO_2$ NWs. The latter signifies strong presence of the in-plane 'O' vacancies over the bridging 'O' vacancies. Similarly, the same ratio was 2:1 for the 25 nm NPs which represented a huge enhancement in bridging 'O' vacancies over the in-plane 'O' vacancies. Among all samples however, the amount of both in-plane and bridging 'O' vacancies were found to be maximum for 25 nm size NPs (Fig. 3(a)).

$CH_4$, a green house and thermodynamically stable gas, was generally detected at high temperature of above 300 °C. Recently it was detected at low temperatures using nanostructured $SnO_2$.[7] This observation was explained following generalized mechanism for metal oxide. In metal oxides, electrons excited from the defect donor states jump to the CB after gaining sufficient energy to embark the conduction process. On adsorption of the atmospheric 'O' species, which vary depending on the operating temperature, a depletion layer of charge is created. This process enhances the resistance of the system and upon exposure to a reducing gas, a decrease in resistance is noticed as a result of releasing back of the adsorbed electrons. In contrast, oxidizing gases increased the resistance for the *n*-type metal oxides.[3] In case of nanostructured $SnO_2$, high response was explained considering its large surface area. However, high response at low temperature was not understood well.[7]

$CH_4$ gas sensing studies were carried out at different temperatures ranging from 50 to 250 °C for all the four samples in the presence of 10% $O_2$. Maximum operating temperature was kept well below the annealing temperature (300 °C) of 4 nm $SnO_2$ NPs to avoid any morphological distortion. Figure 4(a) shows sensor resistance change after exposure of $CH_4$ for all three NPs of 4, 9 and 25 nm at a temperature of 150 °C for the concentration of 500 ppm. The PL data acquired at RT are shown (Fig. 4(b)) for all NPs. The PL intensity was found to vary in similar fashion as the sensor response. This trend was attributed to the presence of increased surface defect density with the increasing size of the NPs as evidenced from the XPS results. Interestingly, sensor response was found to be low for smaller NPs. It indicates that the surface to volume ratio does not play as effective role as surface defects density for sensing at a particular temperature. This observation is in sharp contrast to the earlier report of enhanced sensor response with decreasing crystallite size.[18] Fig. 4 (c) shows the response vs. temperature plots for all of these samples typically for 500 ppm concentration of $CH_4$. Response was defined as $S = (R_{air}-R_{gas})/R_{air}*100$, where $R_{air}$ and $R_{gas}$ were resistances of the system in air and in the exposed gas environment, respectively at constant temperature. For a particular sample, similar response curve with respect to the temperature was observed for all $CH_4$ concentrations (Fig. S5). However nature of response curve with respect to the temperature was not same for all the samples for a particular $CH_4$ concentration (Fig. 4(c)). As mentioned before, response was found to increase with increasing surface defect density of NPs at any temperature for a particular concentration of the exposure gas. Considering effect of operating temperature, the response was found to increase linearly for all the samples until a temperature of 150 °C. The response was found to decrease for 9 nm NPs and $SnO_2$ NWs above 150 °C while 4 nm and 25 nm $SnO_2$ NPs showed enhancement. Such contrasting phenomenon could not be addressed simply by either total amount of defects or by the particle size effect. $CH_4$ gas responses at the lowest measurement temperatures for the respective samples are shown in supplementary information (Fig. S5) Detailed analysis revealed that the in-plane 'O' vacancies played a crucial role in detecting the $CH_4$ at low temperatures. We have also performed a study of selectivity and sensitivity against oxidizing gas oxygen (exposure up to 20%). No significant change in response was observed. Similarly, no sensor response could be detected for reducing agent loke $H_2$ of 200 ppm at 150 °C.

In view of understanding the role of defects and surface reactivity at low temperature, activation energy (AE) for all the samples was calculated using Arrhenius plots of $\ln(S \%)$ vs. $1/T$. This AE being independent of concentrations reflects intrinsic materials property for sensing. For a particular surface reaction on the $SnO_2$ at fixed temperature range, the available free energies will be governed by the characteristics of the surfaces markedly by the defects in nanostructure materials. Arrhenius plots for all samples which were exposed to a typical concentration of 500 ppm of $CH_4$ are shown in (Fig. 4(d)). The 4 nm NPs exhibited AE of 418(15) meV whereas the 9 nm NP and the NWs showed very low AE of 158(10) and 156(15) meV, respectively. In case of the 25 nm NPs two different slopes for two different temperature ranges were observed (Fig. 4d). At low temperature range (until 150 °C) AE was found to be 211(17) meV which was relatively close to the observed AEs of the 9 nm NPs and NWs. For the temperature range at and above of 200 °C, the AE was 440(20) meV, of 25 nm NPs matched with that for the 4 nm NPs. PL analysis also confirmed the presence of different shallow donor sates of SD1, SD2, SD3 and SD4 in the $SnO_2$ nanostructures. However the defect densities pertaining to these states were found to vary with sample size and dimensions. For instance, the 4 nm NPs had high amount of bridging 'O' vacancies (SD3 and SD4) in comparison to the in-plane 'O' vacancies (SD1 and SD2) (Fig. S1a). Donor electrons in SD3 and SD4 levels of the 4 nm NPs needed high energy to reach to the CB for participation in the resistive sensing process. Predictably the corresponding AE for sensing was also found to be high (418(15) meV). Thus, detection was possible only above a temperature 150 °C with reasonable response for the 4 nm NPs. However, its response increased after a temperature of 200 °C. In the case of NWs and the 9 nm NPs, the in-plane 'O' vacancies related luminescence was strongly observed at 2.28 eV (SD2) and 2.45 eV (SD1) (Fig. S1(d)). Since these defect states were close to the CB, it did not require high energy to transmit electrons to the CB. Consequently, both the 9 nm NPs and NWs exhibited low AE and could sense $CH_4$ at low temperatures of 50 °C and 100 °C, respectively (Fig. S5(b and d)). Above 200 °C, both the NWs and the 9 nm $SnO_2$ NPs showed a decrease in the response. Incidentally, low amount of bridging 'O' vacancies (SD3 and SD4) were observed for these samples. In contrast, 25 nm NPs contained both defects in sufficient quantity (Fig. S1(c)) and responded well in both the temperature ranges with two different AE values. One of the AE values of 211(17) meV matched well to the observed values for the 9 nm NPs and NWs. Another value of and 440(20) meV closely followed AE of to 4 nm NPs. Thus the observation indicated that defects corresponding to SD1 and SD2 (in-plane 'O' vacancies) participated at the lower range of temperatures whereas SD3 and SD4 defects (bridging 'O' vacancies) took part at the higher range of temperature in the reducing $CH_4$ gas sensing process.

In conclusion surface defect density, nature of defects and its relative position in band diagram are considered to be important parameters for $CH_4$ sensing process in $SnO_2$ nanostructures. High surface to volume ratio does not influence the sensing substantially without having necessary surface defects. In-plane and bridging 'O' vacancy related defect state positions in the band diagram are explains the temperature dependent $CH_4$ response. In-plane 'O' vacancies creating defect states close to the conduction band minimum are found to be strongly sensitive at low temperature due to low activation energy in comparison to the bridging 'O' vacancies. Such effects may also be observed for other oxides with similar structure.

We thank Dr. S. Amirthapandian, MPD, MSG, IGCAR for his help in performing TEM studies. We acknowledge Syamal Rao Polaki, SND, MSG, IGCAR for his help in recording SEM images.

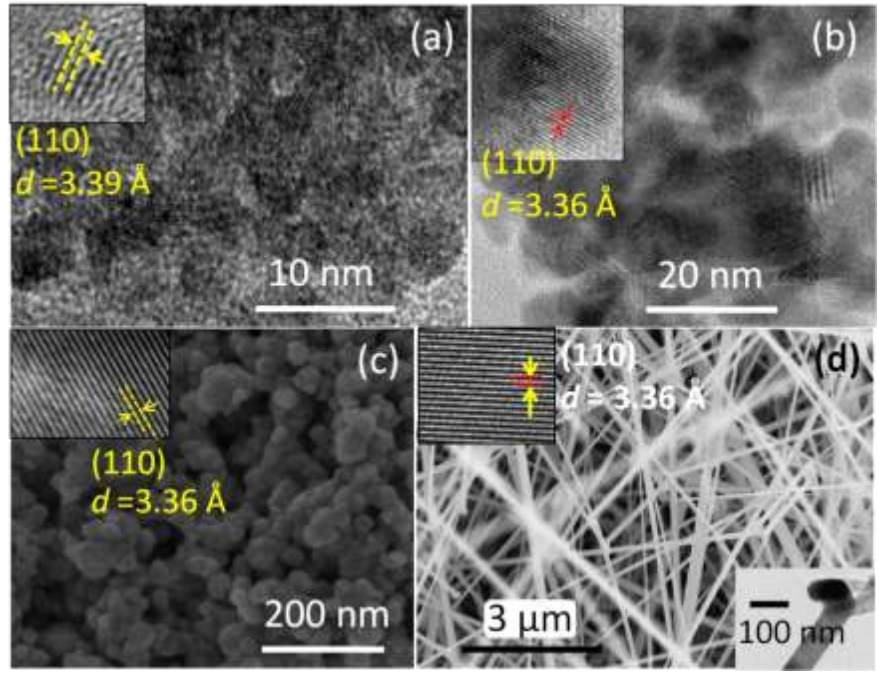

FIG. 1 TEM image of $SnO_2$ NPs annealed at (a) 300 °C (b) 500 °C, FESEM images of (c) $SnO_2$ NPs a annealed at 800 °C and (d) CVD grown $SnO_2$ NWs. Insets shows zoomed HRTEM images.

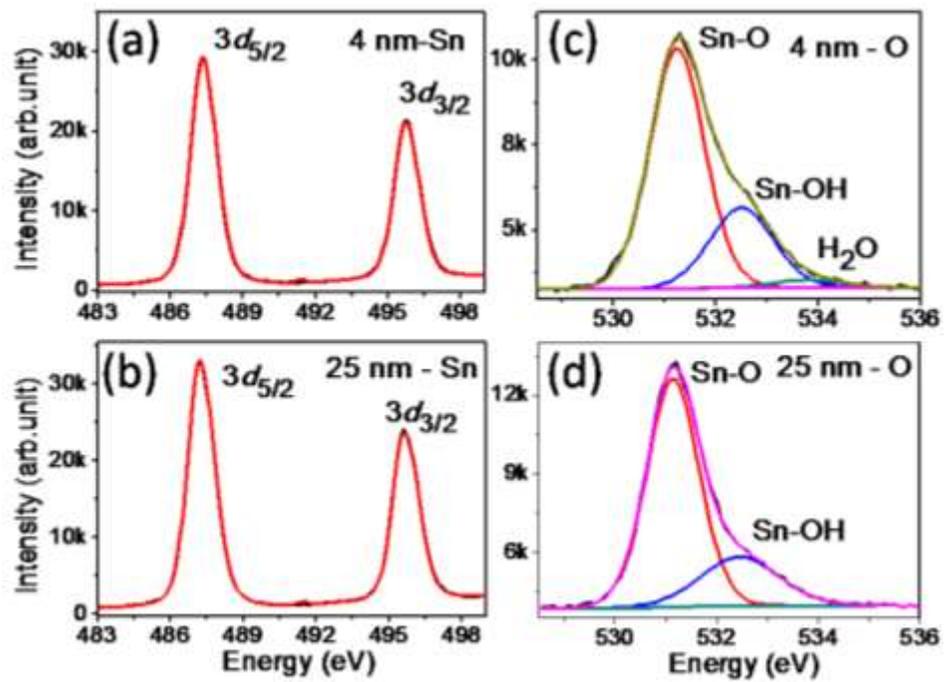

FIG. 2 XPS spectra of NPs (a1) Sn 3d (a2) O 1s of $SnO_2$ (4 nm) and (b1) Sn 3d (b2) O 1s of $SnO_2$ (25 nm).

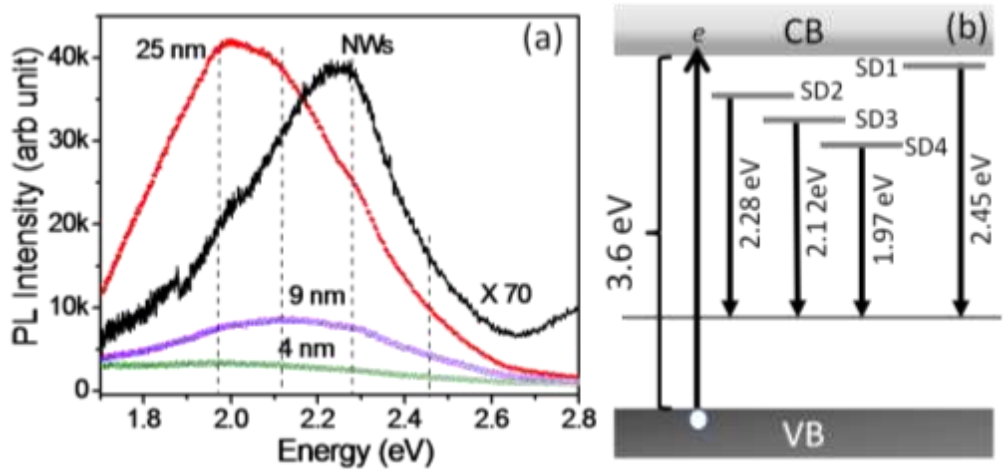

FIG. 3 (a) RT PL spectra of SnO$_2$ NPs and NWs (b) Proposed band diagram

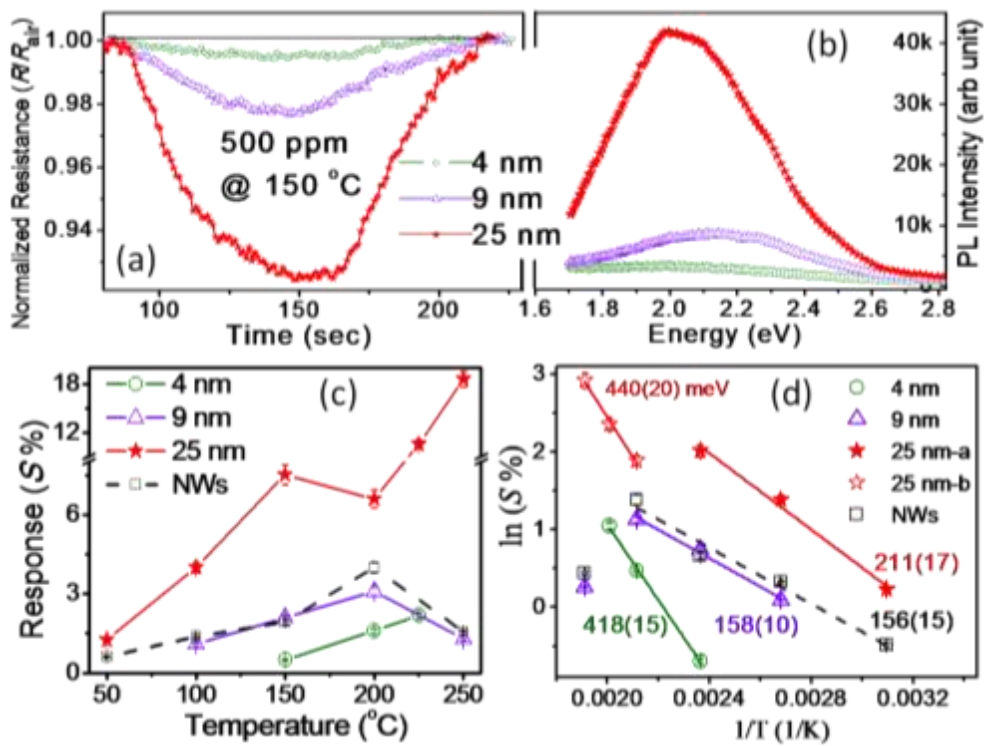

FIG. 4 (a) CH$_4$ normalized resistance and (b) RT PL of SnO$_2$ NPs (c) Gas response of SnO$_2$ NPs and NWs for 500 ppm of CH$_4$ with respect to temperature. (d) Arrhenius plots for NPs and NWs and corresponding AE values for 500 ppm of CH$_4$.